\documentclass[aps,preprint,a4paper,showpacs,showkeys,superscriptaddress]{revtex4-1}

\usepackage{latexsym}
\usepackage{amsmath,amssymb}
\usepackage{graphicx}
\usepackage{subfigure}
\usepackage{xcolor}
\usepackage{cancel}
\usepackage{multirow,tabularx,boldline,array}

\newcolumntype{C}[1]{>{\centering\arraybackslash}p{#1}}
\usepackage{hyperref}     
\hypersetup{colorlinks,%
  citecolor=blue,%
  linkcolor=cyan,%
  pdftex}

\usepackage[titletoc]{appendix}
\usepackage{enumerate}

\begin{document}


\title{Formation of the Hayward black hole from a collapsing shell}

\author{Hwajin Um}%
\email[]{um16@sogang.ac.kr}%
\affiliation{Department of Physics, Sogang University, Seoul, 04107,
  Republic of Korea}%

\author{Wontae Kim}%
\email[]{wtkim@sogang.ac.kr}%
\affiliation{Department of Physics, Sogang University, Seoul, 04107,
  Republic of Korea}%

\date{\today}

\begin{abstract}
  We consider a collapsing shell of matter to form the Hayward black hole and investigate semi-classically
  quantum radiation from the shell.
  Using the Israel's formulation, we obtain the mass relation between the collapsing shell and the Hayward black hole.
  By using the functional Schr\"odinger formulation for the massless quantum radiation,
  the evolution of a vacuum state for a scalar field is shown to be unitary.
  We find that
  the number of quanta at a low frequency decreases for a large length parameter characterizing
  the Hayward black hole.
  Moreover, in the limit of low frequency, the Hawking temperature can be read off from the occupation number of excited states
  when the shell approaches its own horizon.
\end{abstract}

%



\maketitle


\newpage
\section{introduction}
\label{sec:introduction}
Since physical singularities of black holes have been interpreted as a breakdown of general relativity,
there has been much attention to singularity-free black holes, the so-called regular black holes
\cite{Bardeen:1968,Borde:1996df,AyonBeato:1999ec,Hayward:2005gi,Bambi:2013ufa,Balart:2014cga}.
Among these black hole solutions, in particular, the metric of the Hayward black hole
reduces to the geometry of Schwarzschild black hole for a large radius, and to the metric of
the de Sitter spacetime for a small radius so that the curvature becomes nonsingular at the center \cite{Hayward:2005gi}.
In the Hayward black hole, various aspects have been also studied in Refs. \cite{Halilsoy:2013iza,Abbas:2014oua,Amir:2015pja,Pourhassan:2016qoz,Chiba:2017nml,Mehdipour:2016vxh}.

On the other hand,
the functional Schr\"odinger picture
was formulated in order for studying quantum radiation from a collapsing shell
\cite{Vachaspati:2006ki,Vachaspati:2007hr}.
The process of the collapsing shell to form the Schwarzschild black hole was constructed, so
massless quantum radiation as seen by an observer at asymptotic infinity turned out to be nonthermal.
Subsequently, there have been several works employing this method such as
studies on radiation as seen by an infalling observer \cite{Greenwood:2008zg},
collapsing shells to form
the Ba\~{n}ados-Teitelboim-Zanelli black string \cite{Greenwood:2009gp}, Hawking radiation from the Reissner-Nordstr\"om domain wall
\cite{Greenwood:2009pd}, and a massive quantum radiation \cite{Greenwood:2010sx}.
In addition, thanks to an analytic solution of the functional Sch\"odinger equation in Ref.~\cite{Kolopanis:2013sty},
the density matrix of the quantum radiation was also calculated explicitly,
which shows that the process of the radiation is unitary during the evolution
\cite{Saini:2015dea,Saini:2016rrt}.
This fact was confirmed in the other singular black holes such as
the anti-de Sitter Schwarzschild black hole and the Reissner-Nordstr\"om black hole
\cite{Saini:2017mur,Das:2019iru}.
Now, one might wonder how this collapsing process including the functional Schrodinger method works in
the Hayward black hole as a regular black hole.

In this paper, we will study the process of a collapsing shell to form the Hayward black hole and
investigate the quantum radiation in the context of the functional Schr\"odinger equation.
In Sec.~\ref{sec:classical},
we will obtain a mass relation between the collapsing shell and the Hayward black hole by using the Israel formulation \cite{Israel:1966rt}.
Next, in Sec.~\ref{sec:quantum}, a massless quantum radiation from the shell will be studied
by using the functional Schr\"odinger equation
for two cases: when the shell forms the nonextremal black hole and when the horizon is absent.
In the former case, the analytic solution of the functional Schr\"odinger equation will be obtained in the Hayward black hole along the line of Refs.~\cite{Vachaspati:2006ki,Kolopanis:2013sty},
and eventually, the process is shown to be unitary.
In the latter case, we will take the limit of a small radius for
the final stage of the collapsing shell,
and thus, the analytic solution will also be obtained.
In Sec.~\ref{sec:occupation}, in the two cases in Sec.~\ref{sec:quantum}, we will calculate the occupation number of excited states
and discuss the spectrum of the occupation number.
In the former case, the occupation number at a low frequency will be shown to decrease for a larger value of $\ell$
being a length parameter in the Hayward black hole.
In the latter case, the spectrum of the occupation number will turn out to be damped oscillating for the final stage of the collapsing.
Finally, a conclusion and discussion will be given in Sec.~\ref{sec:conclusion}.

\section{classical theory for a collapsing shell to form the Hayward black hole}
\label{sec:classical}

The spacetime will be divided into
the interior and the exterior regions of an infinitely thin shell with a radius $R(t)$ \cite{Vachaspati:2006ki,Vachaspati:2007hr,Greenwood:2008zg,Greenwood:2009gp,Greenwood:2009pd,Greenwood:2010sx,Saini:2015dea,Saini:2016rrt,Saini:2017mur,Das:2019iru}. In the exterior region of the shell, i.e., $r>R(t)$, the spacetime is described by the Hayward metric \cite{Hayward:2005gi},
    \begin{equation}
    \label{eq:exterior}
    \left(ds_+ \right)^2 = -f(r) dt^2+\frac{1}{f(r)} dr^2 +r^2 d\Omega^2
    \end{equation}
with $f(r) = 1-2Mr^2/(r^3+2 \ell^2 M)$,
where $M$ is the mass of the Hayward black hole and $\ell$ is the length parameter
related to cosmological constant as $\Lambda=3/\ell^2$.
The metric \eqref{eq:exterior} reduces to the Schwarzschild
metric for $\ell=0$, and it becomes Minkowski spacetime for $M=0$.
Note that there is a critical mass $M^* = 3\sqrt 3 \ell /4$
to form a black hole such that $\ell_{\max}=4 M^*/(3\sqrt 3)$ \cite{Hayward:2005gi};
$f(r)$ has two simple zeros if $M>M^*$ ($\ell<\ell_{\max}$),
one double zero if $M=M^*$ ($\ell=\ell_{\max}$),
and no zeros if $M<M^*$ ($\ell>\ell_{\max}$).
Next, in the interior region of the shell, the spacetime is assumed to be Minkowski spacetime described by the metric,
	\begin{equation}
    \label{eq:interior}
	\left(ds_- \right)^2 = -dT^2 + dr^2 +r^2 d\Omega^2,
	\end{equation}
which is valid for $r<R(t)$.
On the shell $r=R(t)$,
the proper distance is required to be continuous as $\left(ds_+\right)^2 |_{r\to R(t)} =  \left(ds_-\right)^2 |_{r\to R(t)}$, so a relation between the exterior and interior times is obtained as
	\begin{equation}
	\label{eq:time-rel}
	\frac{dT}{dt} \bigg|_{r=R(t)}= \sqrt{ f(R)+\left(1-\frac{1}{f(R)}\right) \dot{R}^2 },
	\end{equation}
where $\dot{R}=dR/dt$.

In the Israel formulation \cite{Israel:1966rt}, a combined metric can be obtained as
		$g_{\mu\nu} = \Theta(r-R(t)) \tilde{g}^+_{\mu\nu}	+\Theta(-r+R(t)) \tilde{g}^-_{\mu\nu}$,
	where $\Theta(x)$ is the unit step function defined as $\Theta(x) = 1 (x>0)$ and $\Theta(x) =0 (x<0)$
	with the metrics being $\tilde g^+_{\mu\nu}$ and $\tilde g^-_{\mu\nu}$ in
the exterior and the interior regions, respectively.
Then, a junction condition requires that
     $g_{\mu\nu}$ be continuous on the shell,
so the Einstein tensor can be calculated as
	\begin{equation}
    \label{eq:Einstein tensor}
    G_{\mu\nu} = \Theta(r-R(t)) \tilde{G}^+_{\mu\nu}	+\Theta(-r+R(t)) \tilde{G}^-_{\mu\nu} +\delta(r-R(t)) \tilde{G}^0_{\mu\nu},
	\end{equation}
where $\tilde{G}^\pm_{\mu\nu}$ and $\tilde{G}^0_{\mu\nu}$ are Einstein tensors calculated in the exterior and interior regions, and on the shell.
Explicitly,
$\tilde{G}^0_{\mu\nu} = \left\{\kappa_{\mu\gamma} n^\gamma n_\nu \! + \! n_\mu n^\gamma \kappa_{\gamma\nu} \! + \! \kappa_{\gamma}^{\gamma} (g_{\mu\nu} \! - \! n_\mu n_\nu)\! - \!\kappa_{\mu\nu}\! - \!g_{\mu\nu} n^\gamma \kappa_{\gamma \delta} n^\delta \right\}/2$,
where $\kappa_{\mu\nu} = n^{\gamma} \left( \partial_\gamma \tilde g_{\mu\nu}^+ |_{r\to R(t)} - \partial_\gamma \tilde g_{\mu\nu}^- |_{r\to R(t)} \right)$ and $n^\mu$ is a unit normal vector to the shell.
From the Einstein tensor \eqref{eq:Einstein tensor},
the corresponding energy-momentum tensor can also be calculated as
    \begin{equation}
    T_{\mu\nu} = \Theta(r-R(t)) \tilde{T}^+_{\mu\nu}	+\Theta(-r+R(t)) \tilde{T}^-_{\mu\nu} +\delta(r-R(t)) \tilde{T}^0_{\mu\nu}.
	\end{equation}
Projecting $\tilde G^0_{\mu\nu} = 8\pi \tilde T^0_{\mu\nu}$ to the shell
by using the projection operator $h_{\mu\nu} = g_{\mu\nu} - n_{\mu} n_{\nu}$,
we obtain the so-called Israel relation,
	\begin{equation}
	\label{eq:israel}
	 \left[ K h_{\mu \nu} \right]-\left[ K_{\mu \nu} \right] = 8\pi S_{\mu\nu},
	\end{equation}
where $K_{\mu\nu}$ is the extrinsic curvature of the shell and $[...]$ denotes
$[A] = \lim_{r\to R+} A(r) - \lim_{r\to R-} A(r)$.
Now, the source on the shell is assumed to be described by a perfect fluid; thus,
$S^{\mu\nu} =  h_\mu^\gamma \tilde T_{\gamma\nu}^0 = (\sigma-\tau) u^\mu u^\nu -\tau h^{\mu\nu}$,
where $\sigma$ is an energy density and $\tau$ is a surface tension \cite{Israel:1966rt}.
In particular, a domain wall satisfying the relation $\sigma=\tau$ can be chosen as a source, so
the source on the shell reduces to $S^{\mu\nu} = -\sigma h^{\mu\nu}$.
Note that the energy density turns out to be constant thanks to the conservation equation on the shell $h_{\mu \gamma} D_{\nu} S^{\gamma\nu} =0$, where $D_\mu = h_\mu^\nu \nabla_\nu$
\cite{Ipser:1983db,Lopez:1988gt}.

Plugging $S_{\mu\nu}$ into Eq.~\eqref{eq:israel}, we finally obtain the mass relation,
    \begin{equation}
    \label{eq:mass rel}
    \frac{f(R)}{\sqrt{f(R)-\frac{\dot R^2}{f(R)}}} - \frac{1}{\sqrt{1-\frac{\dot R^2}{B^2}}} = -4\pi\sigma R,
    \end{equation}
with $B^2 =(dT/dt)^2 \big|_{r=R(t)}$.
For a static limit, i.e., ${\dot R}=0$,
the mass in Eq.~\eqref{eq:exterior} can be expressed in terms of
the mass $M_0=4\pi R^2\sigma$ on the shell in such a way that
\begin{equation}
M=\frac{M_0-\frac{M_0^2}{2 R}}{1-\frac{2\ell^2}{R^3}\left(M_0-\frac{M_0^2}{2 R}\right)},
\end{equation}
which can be reduced to the case of the Schwarzschild black hole as $\ell\to 0$ \cite{Vachaspati:2006ki}.
From Eq.~\eqref{eq:mass rel} with Eq.~\eqref{eq:time-rel},
a first-order differential equation can be obtained as
	\begin{equation}
	\label{eq:dR over dt}
	\frac{dR}{dt}=-\frac{f(R)}{ \sqrt{1+f(R)\left[1-\frac{R^2}{4M_0^2} \left ( M_0^2+1-f(R) \right )^2\right]^{-1}}}.
	\end{equation}

In order to study the collapsing shell to form the Hayward black hole,
we will investigate the case of $M>M^*$.
When the collapsing shell starts to form the black hole as $R(t) \to R_H$,
Eq.~\eqref{eq:dR over dt} reduces to a simple form as \cite{Vachaspati:2006ki,Vachaspati:2007hr,Greenwood:2008zg,Greenwood:2009gp,Greenwood:2009pd,Greenwood:2010sx,Saini:2015dea,Saini:2016rrt,Saini:2017mur,Das:2019iru}
	\begin{equation}
    \label{eq:R_limit}
	\frac{dR}{dt} \approx -f(R).	
	\end{equation}
Solving this equation in this {\it incipient limit}, we get the radius
$R(t) = R_H + (R_0 -R_H) e^{-f'(R_H)t}$ for $0\leq t \leq t_f$,
while the initial and the final radii are $R_0=R(t_0)$ for $t<0$ and $R_f=R(t_f)$ for $t>t_f$, respectively.

In the case of $M<M^*$ where the horizon is absent,
$(dR/dt)$ is regular,
the shell shrinks continuously and then finally reaches $R=0$.
Since it would be a highly nontrivial task to solve Eq.~\eqref{eq:dR over dt} analytically in the whole region,
we will discuss the limiting case of a small radius ($R\to 0$).
As $R\to 0$ in Eq.~\eqref{eq:dR over dt},
$(dR/dt)\to -1/\sqrt 2$ and so,
\begin{equation}
\label{eq:asymptotic R(t) for a small}
    R(t)=R_0 - \frac{t}{\sqrt 2},
\end{equation}
where $t_{\rm max} = \sqrt 2 R_0$ for the positivity of the radius.

\section{quantum radiation from the collapsing shell}
\label{sec:quantum}
If an observer at asymptotic infinity sees the formation of the black hole,
one can ask what radiation that characterizes gravitational collapse might be observed.
Hence, we consider a quantum scalar field on the background of the collapsing shell
and derive a quantized theory of the scalar field on the Hayward black hole.

The total action for a single scalar field is assumed to consist of both the
exterior and the interior actions such as
    \begin{eqnarray}
    \label{totalaction}
    S_\Phi &=& S^+_\Phi +S^-_\Phi, \\
    S^\pm_\Phi &=& \int {\textrm d}^4 x_\pm \sqrt{-g^\pm} \left(-\frac 1 2 g^\pm ~^{\mu\nu} \partial^\pm_{\mu} \Phi \partial^\pm_{\nu} \Phi \right), \label{eachactions}
	\end{eqnarray}
where $g^+_{\mu\nu}$ and $g^-_{\mu\nu}$ are the Hayward metric \eqref{eq:exterior} and the Minkowski metric \eqref{eq:interior}, and $x_{\pm}^\mu$ are the exterior and the interior coordinates, respectively.
By using the spherical symmetries of the metrics,
the scalar field can be decomposed into real spherical harmonics as
	\begin{equation}
    \label{eq:spherical symm sol}
	\Phi(x_\pm^0,r,\theta,\phi)=\sum_{j, m_j } \Phi_j (x_\pm^0,r) Y_{j,m_j} (\theta,\phi)
	\end{equation}
with $j=0,1,2,...$ and $m_j = -j,-j+1,...,j$.
Then, the actions \eqref{eachactions} can be written as
	\begin{align}
    \label{eq:action outside}
	S^+_\Phi & = \sum_{j} 4\pi (2j+1) \int \!\!{\textrm d} t \int_R^\infty \!\!{\textrm d} r
	\left( \frac{r^2}{2 f(r)}  \left(\frac{\partial \Phi_j}{\partial t} \right)^2
	-\frac{r^2 f(r)}{2} \left(\frac{\partial \Phi_j}{\partial r} \right)^2 - \frac{j(j+1)}{2} \Phi_j^2 \right), \\ \label{eq:action inside}
	S^-_\Phi & = \sum_{j} 4\pi (2j+1) \int {\textrm d} T \int_0^R {\textrm d} r
	\left( \frac{r^2}{2}  \left(\frac{\partial \Phi_j}{\partial T} \right)^2
	-\frac{r^2}{2} \left(\frac{\partial \Phi_j}{\partial r} \right)^2 - \frac{j(j+1)}{2} \Phi_j^2 \right).
	\end{align}
Especially, the interior action \eqref{eq:action inside} is written as
	\begin{equation}
	S^-_\Phi = \sum_{j} 4\pi (2j+1) \int {\textrm d} t \int_0^R {\textrm d} r
	\left( \frac{r^2}{2B}  \left(\frac{\partial \Phi_j}{\partial t} \right)^2
	- \frac{r^2 B}{2} \left(\frac{\partial \Phi_j}{\partial r} \right)^2 - \frac{j(j+1) B}{2} \Phi_j^2 \right),
\label{eq:action inside w B}
	\end{equation}
where we have used the relation $B = \left(dT/dt\right)|_{r=R(t)}$ in Eq.~\eqref{eq:time-rel}.
In the functional Schr\"odinger formalism, one can derive the Schr\"odinger equation
by using the Hamiltonian given as the Legendre transformation of the action \eqref{totalaction}.
Since $R(t)$ was calculated analytically for the cases of $M>M^*$ and $M<M^*$ in the previous section
and $B$ in Eq.~\eqref{eq:action inside w B} is a function of $R(t)$,
we can study the Hamiltonian analytically in each case.
In the following subsections,
Secs.~\ref{sec:quantum_blackhole} and \ref{sec:quantum_no_blackhole},
we will study the quantum radiation by using the functional Schr\"odinger formalism for $M>M^*$ and $M<M^*$.

\subsection{Presence of the horizon ($M>M^*$)}
\label{sec:quantum_blackhole}
In order to study a nonextremal black hole, the case of $M>M^*$ will be considered.
We decompose $\Phi_j$
into a complete set of real basis functions denoted by $\{d_{k,j}\}$ \cite{Vachaspati:2006ki},
	\begin{equation}
	\label{eq:mode sum}
	\Phi_j = \sum_k a_{k,j}(t) d_{k,j} (r),
	\end{equation}
where $a_{k,j}(t)$ is a mode amplitude.
By using Eq.~\eqref{eq:mode sum}, in the incipient limit $R \to R_H$,
the action \eqref{totalaction} is obtained as
	\begin{equation}
    \label{eq:action approx}
	S_\Phi \approx 	\sum_{k,k',j} (2j+1) \int {\textrm d} t \left(\frac{1}{2f(R)} \dot a _{k,j} \tilde A_{k,k',j} \dot a _{k',j}
	- \frac 1 2 a_{k,j} (\tilde B_{k,k',j} + \tilde C_{k,k',j}) a_{k',j} \right)
	\end{equation}
after spatial integrations, where the coefficients are defined by
\begin{align}
\tilde A_{k,k',j} & =4\pi \int_0^{R_H} {\textrm d}r r^2 d_{k,j} d_{k',j}~,\\
\tilde B_{k,k',j} & =4\pi \int_{R_H}^\infty {\textrm d} r r^2 f(r) \frac{{\textrm d} d_{k,j}}{{\textrm d} r} \frac{{\textrm d} d_{k',j}}{{\textrm d}r}~,\\
\tilde C_{k,k',j} & =4\pi j(j+1) \int_{R_H}^\infty{\textrm d}r d_{k,j} d_{k',j}.
\end{align}
Since the coefficients $\tilde A_{k,k',j}$, $\tilde B_{k,k',j}$, and $\tilde C_{k,k',j}$ are Hermitian operators,
they can be simultaneously diagonalized by a set of eigenvectors $\{b_{k,j}\}$ \cite{goldstein:mechanics,Vachaspati:2006ki},
	\begin{equation}
	\label{eq:additional1}
	S_\Phi =\sum_{k,j} (2j+1) \int {\textrm d} t \left(\frac{1}{2f(R)} \alpha_{k,j} \dot b _{k,j}^2
	- \frac 1 2 (\beta_{k,j} + \gamma_{k,j}) b_{k,j}^2 \right),
	\end{equation}
where $\alpha_{k,j}$, $\beta_{k,j}$, and $\gamma_{k,j}$ are eigenvalues of $\tilde A_{k,k',j}$, $\tilde B_{k,k',j}$, and $\tilde C_{k,k',j}$, respectively.

We are now in a position to impose the quantization rule as $\left[b_{k,j},\Pi_{k',j'} \right] = i \delta_{k,k'} \delta_{j,j'}$
represented by $\Pi_{k,j} \to -i \partial/\partial b_{k,j}$,
where $\Pi_{k,j}=\partial L_\Phi/\partial \dot b_{k,j} =(2j+1) \alpha_{k,j} \dot b_{k,j} / f(R)$. Hence, the Hamiltonian for the quantum radiation can be obtained as
	\begin{equation}
	\label{eq:additional2}
	H_\Phi = \sum_{k,j} \left(-\frac{f(R)}{2 (2j+1) \alpha_{k,j}} \frac{\partial^2}{\partial b_{k,j}^2}
	+ \frac 1 2 (\beta_{k,j} + \gamma_{k,j})  b_{k,j}^2 \right).
	\end{equation}
From the functional Schr\"odinger equation $H_\Phi \Psi= i \partial \Psi /\partial t$ \cite{Vachaspati:2006ki},
the Schr\"odinger equation of the wave function $\psi$ for one eigenvector $b \in\{b_{k,j}\}$ can be written as
	\begin{equation}
    \label{eq:Sch eq}
	\left\{ -\frac{f(R)}{2 \alpha} \frac{\partial^2}{\partial b^2} +\frac 1 2 \alpha \omega_0^2 b^2 \right\} \psi(t,b)
	= i \frac{\partial \psi}{\partial t} (t,b),
	\end{equation}
where $\omega_0 = \sqrt{(2j+1)(\beta+\gamma) / \alpha }$
with $\alpha\in\{(2 j+1)\alpha_{k,j}\}$, $\beta\in\{\beta_{k,j}\}$, and $\gamma\in\{\gamma_{k,j}\}$.
For convenience, a new time parameter is defined as
\begin{equation}
\label{eq:time parameter}
    \eta = \int_0^t\textrm d t'~ f(R(t'))
\end{equation}
so that Eq.~\eqref{eq:Sch eq} becomes the Schr\"{o}dinger equation for a harmonic oscillator with the mass $\alpha$ and the
time-dependent frequency $\omega(\eta)$ as
    \begin{equation}
    \label{eq:Sch eq eta}
    \left[-\frac{1}{2\alpha} \frac{\partial^2}{\partial b^2} +\frac 1 2 \alpha \omega^2 (\eta) b^2 \right]\psi(\eta,b)= i\frac{\partial \psi}{\partial \eta} (\eta, b),
\end{equation}
where $\omega(\eta) = \omega_0/\sqrt{f(R)}$.
For $\eta<0$ and $\eta>\eta_f$, the frequency $\omega(\eta)$ must be constant, so
the eigenstates are given by those of simple harmonic oscillators,
    \begin{equation}
    \label{eq:eigenstate SHO}
    \phi_n (b) =\left( \frac{\alpha \bar \omega}{\pi}\right)^{\frac 1 4} \frac{1}{\sqrt{2^n n!}} H_n (\sqrt{\alpha \bar \omega} b) e^{-\frac 1 2 \alpha \bar \omega b^2},
    \end{equation}
where $H_n (x)$ are Hermite polynomials and $\bar \omega$ is $\omega_0$ for $\eta<0$ or $\omega_f=\omega_0/\sqrt{f(R_f)}$ for $\eta>\eta_f$.

In addition, in the case of $0\leq\eta \leq\eta_f$, the solutions for the time-dependent harmonic oscillator
have been shown to be coherent states, which are the unitary transform of eigenstates \cite{Dantas:1990rk},
and thus,
the solution to Eq.~\eqref{eq:Sch eq eta}
is given by a coherent state with the lowest energy as
    \begin{equation}
    \label{eq:solution}
    \psi(\eta,b)=e^{i \theta( \eta)} \left(\frac{ \alpha}{\pi \zeta^2}\right)^{\frac 1 4} e^{i\left(\frac{1}{\zeta} \frac{\textrm d \zeta}{\textrm d \eta}+\frac{i}{\zeta^2}\right) \frac{\alpha b^2}{2}},
\end{equation}
where $\theta(\eta)=(-1/2)\int_0^{\eta} \textrm d \eta ' \zeta^{-2} (\eta ')$. Note that
$\zeta(\eta)$ is a real solution satisfying
\begin{equation}
\label{eq:diff eq}
    \frac{\textrm d ^2 \zeta}{\textrm d \eta^2}+\frac{\omega_0 ^2}{f(R)} \zeta=\frac{1}{\zeta^3}
\end{equation}
with initial conditions: $\zeta(0)=1/\sqrt{\omega_0}$ and $[\textrm d \zeta/\textrm d \eta] (0)=0$.
In the incipient limit, the metric function is approximately written as
\begin{equation}
\label{eq:asymptotic omega}
    f(R)\approx 1- H \eta,
\end{equation}
where $H=[\textrm d f(R)/\textrm d R] |_{R=R_H}=(R_H^2-3\ell^2)/R_H^3$,
so that the real solution $\zeta(\eta)$ in Eq.~\eqref{eq:diff eq} can be obtained as \cite{Kolopanis:2013sty}
    \begin{equation}
    \label{eq:solution1}
    \zeta(\eta)=\frac{1}{\sqrt{\omega_0}} \sqrt{\xi^2 (\tilde \eta)+\chi^2 (\tilde \eta) },
    \end{equation}
where
    \begin{align}
    \label{eq:solution2}
    \xi (\tilde \eta) &= \frac{\pi y}{2}\left(Y_0 (x) J_1(y)-J_0(x) Y_1(y) \right),\\\label{eq:solution3}
    \chi (\tilde \eta) &= \frac{\pi y}{2}\left(Y_1 (x) J_1(y)-J_1(x) Y_1(y) \right),
    \end{align}
and $J_n$ and $Y_n$ are Bessel functions of the first kind and the second kind, respectively
with $\tilde \eta = H \eta$, $x=2\omega_0/H$, and $y=(2\omega_0/H) \sqrt{1-\tilde \eta}$.
In the incipient limit,
a basis is taken as $\{\phi_n (b) |_{\bar \omega =\omega_f}\}$; and, thus,
the probability amplitude in the $n$th eigenstate in the vacuum state can be calculated as
\begin{align}
    c_{n} (\eta)
    &= \int_{-\infty}^{\infty} \textrm d b~ \phi_n (b)  \psi(\eta,b)\\
    &=\left\{
        \begin{array}{ll}
        e^{i \theta(\eta)} \frac{(n-1)!!}{\sqrt{n!}} \frac{1}{\sqrt[4]{\omega_f \zeta^2 } }
    \sqrt{(-1)^n \frac 2 P \left(1-\frac 2 P\right)^n}  & \qquad (n=0,2,4,...) \\
        0 & \qquad (n=1,3,5,...),\label{eq:prob amp}
        \end{array}
    \right.
\end{align}
where $P=1-(i/\omega_f \zeta) (\textrm d \zeta/\textrm d \eta) + 1/(\omega_f \zeta^2)$.

In connection with the probability amplitude \eqref{eq:prob amp},
the density matrix for $0\leq \eta \leq \eta_f$ takes the form of
$ \rho (\eta,b,b') = \sum_{n,m} c_n (\eta) c_m(\eta) \phi_n (b) \phi_m (b')$ in the incipient limit.
Then the trace of the density matrix is calculated as
    \begin{equation}
    \operatorname{Tr}\rho =\sum_{n=0}^\infty \int\!\!{\textrm d} b \! \int\!\! {\textrm d} b' \phi_n (b) \rho (\eta,b,b') \phi_n (b') = \sum_{n=0}^\infty |c_n (\eta)|^2
    = \sqrt{\frac{H}{\zeta^2 \omega_f}}\frac{2}{|P|}\frac{1}{\sqrt{1-|1-\frac 2 P|^2}}
    \end{equation}
which can be simplified as
    \begin{equation}
    \label{eq:unitarity}
    \operatorname{Tr}\rho =\frac{|\omega_f \zeta^2|}{\sqrt{\omega_f \zeta^2 \left(
    \operatorname{Re}[\omega_f \zeta^2] - \operatorname{Im}[\omega_f \zeta^2] \operatorname{Re}[\zeta \frac{\textrm d \zeta}{\textrm d \eta}]+\operatorname{Re}[\omega_f \zeta]\operatorname{Im}[\zeta \frac{\textrm d \zeta}{\textrm d \eta}] \right)}}=1
    \end{equation}
because $\zeta(\eta)$ is real regardless of $H$.
Therefore, the process of radiation turns out to be unitary during the evolution in the Hayward black hole,
which is similar to the evolution process of singular black holes \cite{Saini:2015dea,Saini:2016rrt,Saini:2017mur,Das:2019iru}.
This fact can also be understood from the conservation law of a probability current.
Using Eq.~\eqref{eq:Sch eq} in the incipient limit, we define the probability and the probability current as
$P_{pro}=|\psi(t,b)|^2$ and
$J (t,b)= (f(R)/(2 \alpha i)) \left(\psi^* \partial_b \psi - \psi\partial_b \psi^* \right)$, respectively,
where they naturally satisfy the continuity equation
$\partial P_{pro} /\partial t +\partial J(t,b) /\partial b =0$.
It leads to $\operatorname{Tr}\rho=1$.

\subsection{Absence of the horizon ($M<M^*$)}
\label{sec:quantum_no_blackhole}

In order to extend our analytical method to $M<M^*$,
we investigate a special case of a small radius of shell.
Because the shell could shrink until $R=0$ at the final stage of the collapsing,
we focus on the limiting case of the small radius in order to what happens finally.

Let us start with the total action \eqref{totalaction} which consists of
the exterior action \eqref{eq:action outside} and the interior action \eqref{eq:action inside w B}.
After decomposing $\Phi_j$ like in Eq.~\eqref{eq:mode sum},
the total action can be written as
\begin{equation}
    \label{eq:total action2}
	S_\Phi =\sum_{k,k',j} (2j+1) \int {\textrm d} t \left(\frac{1}{2} \dot a _{k,j} \tilde A_{k,k',j} (t) \dot a _{k',j}
	- \frac 1 2 a_{k,j} \left(\tilde B_{k,k',j} (t) + \tilde C_{k,k',j} (t)\right) a_{k',j} \right),
\end{equation}
where the coefficients are
\begin{align}
\tilde A_{k,k',j}(t) & = \tilde A_{k,k',j}^+ + \frac{1}{B(t)} \tilde A_{k,k',j}^- ,\\
\tilde B_{k,k',j} (t) & = \tilde B_{k,k',j}^+ + B(t) \tilde B_{k,k',j}^- ,\\
\tilde C_{k,k',j} (t) &= \tilde C_{k,k',j}^+ + B(t) \tilde C_{k,k',j}^- ,
\end{align}
with $B(t) =(dT/dt)|_{r=R(t)}$.
Note that $\tilde A_{k,k',j}^\pm$, $\tilde B_{k,k',j}^\pm$, and $\tilde C_{k,k',j}^\pm$ are time-independent quantities defined by
\begin{align}\label{eq:time-ind coefficients}
\begin{array}{ll}
\tilde A_{k,k',j}^+ = 4\pi \int_{R_0}^{\infty} {\textrm d}r \frac{r^2}{f(r)} d_{k,j} d_{k',j},
& \tilde A_{k,k',j}^-  = 4\pi \int_0^{R_0} {\textrm d}r r^2  d_{k,j} d_{k',j}, \\
\tilde B_{k,k',j}^+ = 4\pi \int_{R_0}^\infty {\textrm d} r r^2 f(r)  \frac{{\textrm d} d_{k,j}}{{\textrm d} r} \frac{{\textrm d} d_{k',j}}{{\textrm d}r},
& \tilde B_{k,k',j}^- = 4\pi \int_{0}^{R_0} {\textrm d} r r^2 \frac{{\textrm d} d_{k,j}}{{\textrm d} r} \frac{{\textrm d} d_{k',j}}{{\textrm d}r},\\
\tilde C_{k,k',j}^+ = 4\pi j(j+1) \int_{R_0}^\infty{\textrm d}r d_{k,j} d_{k',j},
& \tilde C_{k,k',j}^- = 4\pi j(j+1) \int_{0}^{R_0}{\textrm d}r d_{k,j} d_{k',j}.
\end{array}
\end{align}
After diagonalizing the action \eqref{eq:total action2}
\cite{goldstein:mechanics,Vachaspati:2006ki}
and then imposing the quantization rule,
we can obtain the quantized Hamiltonian as
\begin{equation}
	H_\Phi = \sum_{k,j} \left(-\frac{1}{2 (2j+1) \left( \alpha_{k,j}^+ + B(t)^{-1} \alpha_{k,j}^-\right) } \frac{\partial^2}{\partial b_{k,j}^2}
	+ \frac 1 2 (2j+1) \left(\delta_{k,j}^+ + B(t) \delta_{k,j}^-\right)  b_{k,j}^2 \right),
\end{equation}
where $\{b_{k,j}\}$ is a set of eigenvectors,
and $\alpha_{k,j}^\pm$, $\beta_{k,j}^\pm$, and $\gamma_{k,j}^\pm$ are
eigenvalues of the coefficients
$\tilde A_{k,j}^\pm$, $\tilde B_{k,j}^\pm$, and $\tilde C_{k,j}^\pm$
with
$\delta_{k,j}^\pm =\beta_{k,j}^\pm + \gamma_{k,j}^\pm$.

We are going to introduce the functional Schr\"odinger equation
$H_\Phi \Psi= i \partial \Psi /\partial t$ \cite{Vachaspati:2006ki}.
Then, for one eigenvector $b \in\{b_{k,j}\}$,
the equation is explicitly given as
\begin{equation}
\label{eq:HO w time-dep mass and freq}
    \left\{ -\frac{1}{2 (\alpha^+ + B(t)^{-1} \alpha^-)} \frac{\partial^2}{\partial b^2} +\frac 1 2 \left( \delta^+ +B(t) \delta^- \right) b^2 \right\} \psi(t,b)
	= i \frac{\partial \psi}{\partial t} (t,b),
\end{equation}
where $\alpha^\pm \in\{(2 j+1)\alpha_{k,j}^\pm \}$ and $\delta^\pm \in\{(2j+1) \delta_{k,j}^\pm \}$.
Note that this is the Schr\"odinger equation for a harmonic oscillator
with time-dependent mass and frequency.
Thanks to the transformation rule in Ref.~\cite{Dantas:1990rk},
we can rewrite Eq.~\eqref{eq:HO w time-dep mass and freq}
in terms of a time-independent mass as
\begin{equation}
\label{eq:Sch eq in M>M*}
    \left\{-\frac{1}{2 \bar\alpha} \frac{\partial^2}{\partial b^2} + \frac 1 2 \bar\alpha \bar \Omega^2(t) b^2\right\}
    \psi(t,b) = i\frac{\partial\psi}{\partial t} (t,b),
\end{equation}
where $\bar \alpha$ is a constant and
a time-dependent frequency is defined by
\begin{equation}
    \bar \Omega^2 (t) = \frac{\delta^+ +B(t)\delta^-}{\alpha^+ +B(t)^{-1}\alpha^-}
    -\frac 1 4 \left\{\frac{{\textrm d}}{{\textrm d} t} \log \left(\alpha^+ +B(t)^{-1} \alpha^-\right) \right\}^2
    - \frac 1 2 \frac{{\textrm d}^2}{{\textrm d}t^2}
    \log \left(\alpha^+ +B(t)^{-1} \alpha^-\right).
\end{equation}
Since Eq.~\eqref{eq:Sch eq in M>M*} takes the same form with Eq.~\eqref{eq:Sch eq eta} with a different mass and frequency,
the solution to Eq.~\eqref{eq:Sch eq in M>M*} can be easily written as
\begin{equation}
\label{eq:sol in M<M*}
    \psi(t,b) = e^{i\bar \theta(t)} \left(\frac{\bar \alpha}{\pi \bar \zeta^2}\right)^{\frac 1 4} e^{i\left(\frac{1}{\bar \zeta} \frac{\textrm d \bar \zeta}{\textrm d t}+\frac{i}{\bar \zeta^2}\right) \frac{\bar \alpha b^2}{2}},
\end{equation}
where
$\bar \theta(t)=(-1/2)\int_0^{t} \textrm d t' \bar \zeta^{-2} (t')$, and
$\bar \zeta(t)$ is a real solution of
\begin{equation}
\label{eq:zeta in M<M*}
    \frac{\textrm d ^2 \bar \zeta}{\textrm d t^2} + \bar \Omega^2 (t)\bar \zeta=\frac{1}{\bar \zeta^{3}}
\end{equation}
with initial conditions: $\bar \zeta(0)=1/\sqrt{\bar\Omega(0)}$ and $[\textrm d \bar \zeta/\textrm d t] (0)=0$.
\newpage

In order to obtain the analytic solution $\zeta(\eta)$ to Eq.~\eqref{eq:diff eq},
we used the form \eqref{eq:asymptotic omega} in the incipient limit; likewise, a simpler form of $\bar \Omega^2(t)$ will be used for the small radius.
After some calculations, we can get the asymptotic form $\bar\Omega^2 (t) \approx \tilde a_0 +\tilde a_1 R^2$,
where the constants $\tilde a_0$ and $\tilde a_1$ depend on $\alpha^\pm$, $\delta^\pm$, and $\ell$.
Substituting the asymptotic solution \eqref{eq:asymptotic R(t) for a small} into $\bar \Omega^2 (t)$, we can obtain
\begin{equation}
\label{eq:asymptotic Omega}
    \bar\Omega(t)^2 = \tilde b_0 + \tilde b_1 t + \mathcal O (t^2),
\end{equation}
where the constants $\tilde b_0$ and $\tilde b_1$ depend on $\alpha^\pm$, $\delta^\pm$, $\ell$, and $R_0$.
For a small $t$, Eq.~\eqref{eq:asymptotic Omega} can be approximately rewritten as
\begin{equation}
\label{eq:asymptotic omega for t}
    \bar\Omega^2(t) \approx \frac{\tilde b_0}{1-\frac{\tilde b_1}{\tilde b_0} t},
\end{equation}
where the numerator and the denominator in Eq.~\eqref{eq:asymptotic omega for t} correspond to $\omega_0^2$ in Eq.~\eqref{eq:diff eq} and
$f(R)$ in Eq.~\eqref{eq:asymptotic omega}, respectively.
By using Eqs.~\eqref{eq:solution1},\eqref{eq:solution2}, and \eqref{eq:solution3},
we can finally obtain the coherent state \eqref{eq:sol in M<M*}.

It is worth noting that $\bar \Omega^2(t)>0$ in Eq.~\eqref{eq:asymptotic omega for t} so that $\tilde b_0 >0$
since $\bar \Omega^2(t)$ is positive even for $t=0$.
Hence, $t<\tilde b_0 / \tilde b_1$.
On the other hand, from the asymptotic solution \eqref{eq:asymptotic R(t) for a small},
$t_{\rm max} = \sqrt 2 R_0$.
If we require  $t<\min ( \tilde b_0 / \tilde b_1,\sqrt 2 R_0)$,
the analytic solution \eqref{eq:sol in M<M*} can be well-defined for $0\leq t<\min ( \tilde b_0 / \tilde b_1,\sqrt 2 R_0)$.
Now, we will study the special case of $\tilde b_0 / \tilde b_1 > \sqrt 2 R_0$ for simplicity,
so that the analytic solution \eqref{eq:sol in M<M*} can be well-defined
during the collapsing from the initial radius $R_0$ to the final zero radius where $0 \leq t < t_{\rm max}$.

\section{Occupation number and Hawking temperature}
\label{sec:occupation}

If we consider detectors at asymptotic infinity designed to register particles for the quantum radiation,
the number of quanta is obtained by evaluating
the occupation number of the excited states
\cite{Vachaspati:2006ki},
    \begin{equation}
    \label{eq:occ num def}
    N\left( \eta,\omega \right)
    =\sum_{n=0}^{\infty} n |c_n (\eta) |^2,
    \end{equation}
where $c_n(\eta)$ is the probability amplitude in the $n$th excited state \eqref{eq:prob amp}.
From the spectrum of the occupation number \eqref{eq:occ num def},
the radiation of the collapsing shell to form a singular black hole
at early times turned out to be nonthermal
\cite{Vachaspati:2006ki,Vachaspati:2007hr,Greenwood:2008zg,Greenwood:2009pd,Kolopanis:2013sty,Saini:2015dea,Saini:2016rrt,Saini:2017mur};
however, the spectrum
interestingly resembles the thermal Hawking distribution when the shell approaches its own horizon.

\begin{figure}[b]
\label{fig:occ num}
\centering
\subfigure[]{\includegraphics[width=0.49\textwidth]{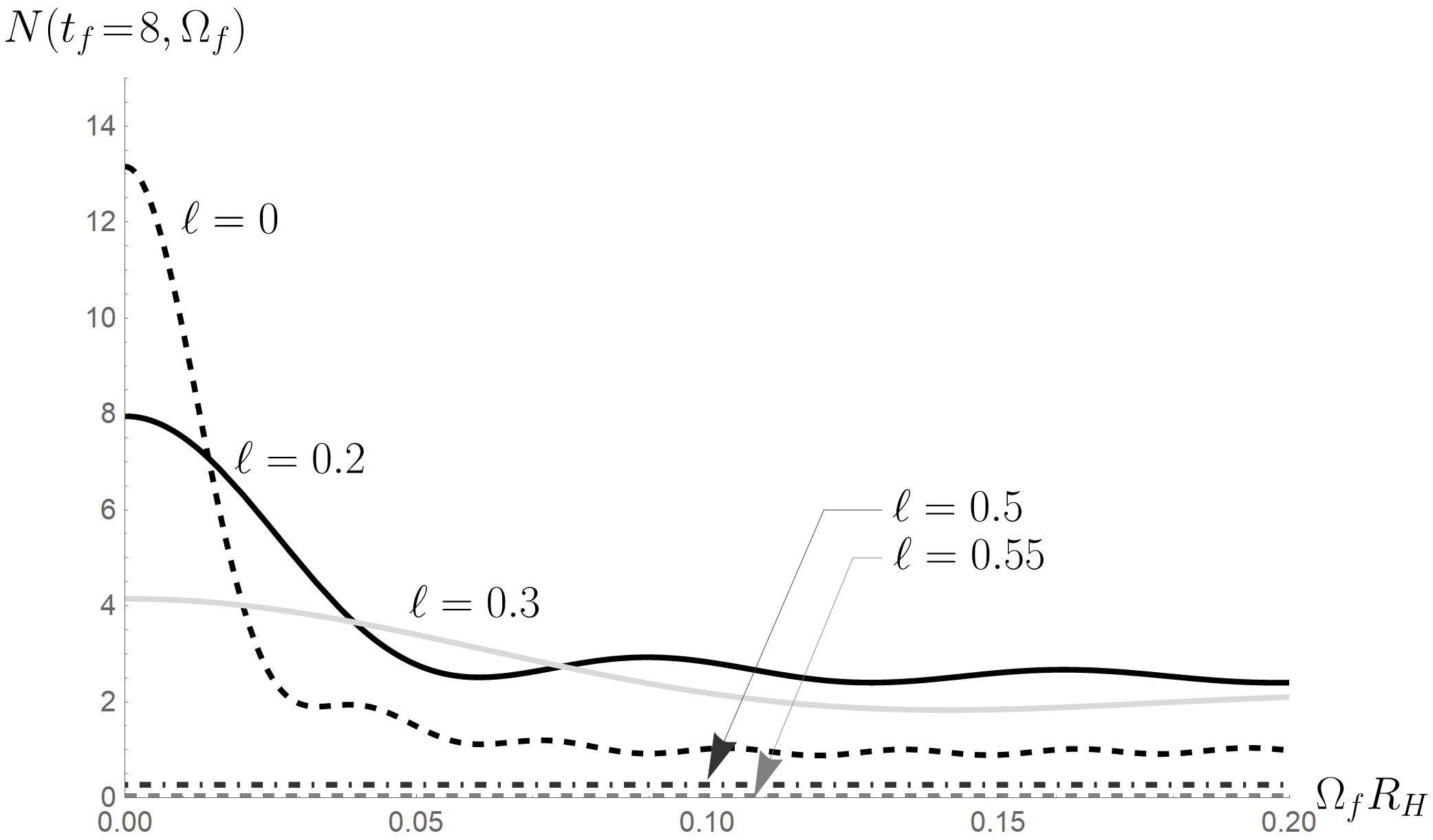}\label{fig:a}}
\subfigure[]{\includegraphics[width=0.49\textwidth]{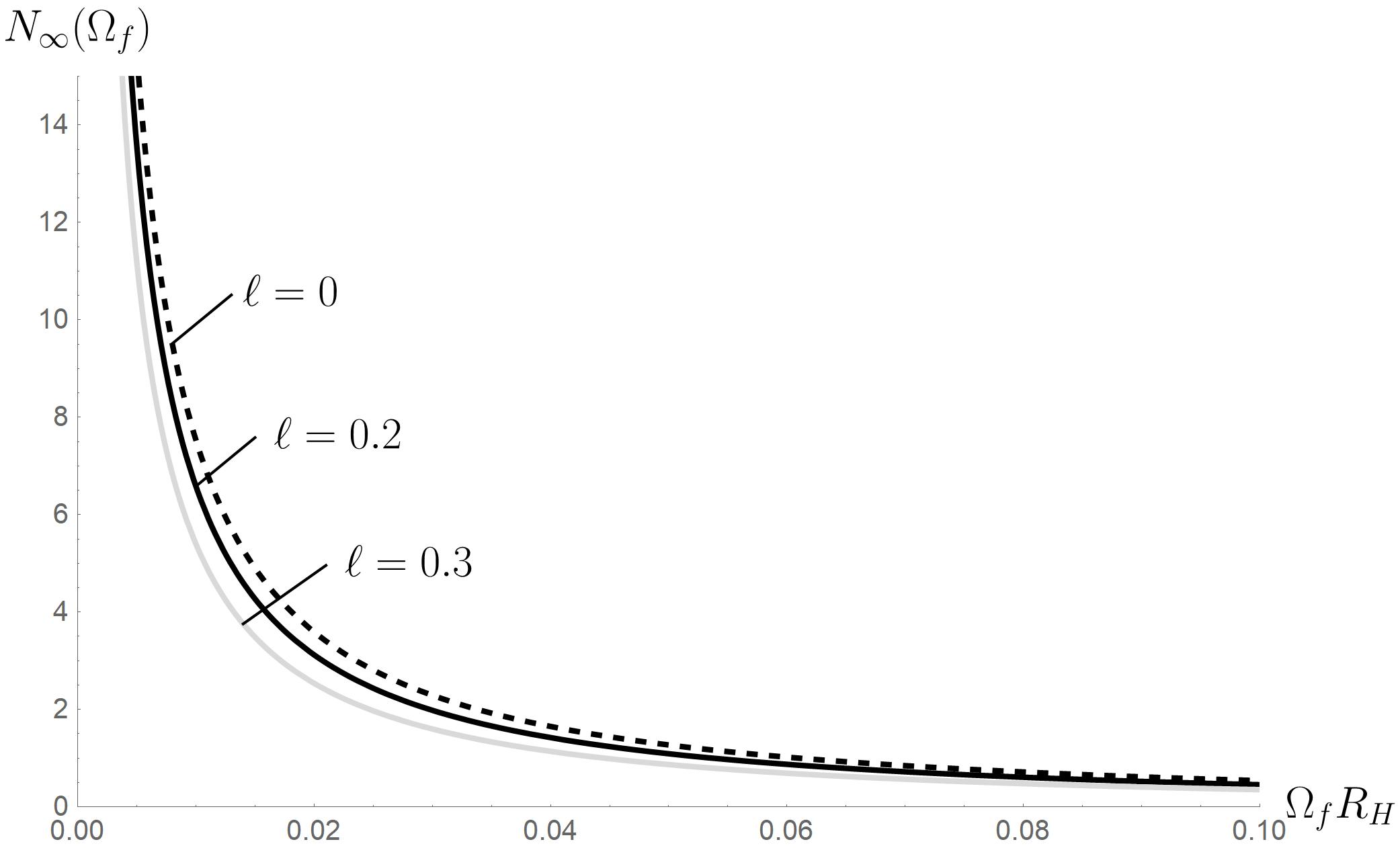}\label{fig:b}}
\caption{The occupation number \eqref{eq:finite time} for a finite time is plotted in Fig.~1(a). Figure 1(b) also shows the spectrum of Eq.~\eqref{eq:infinite time}
which is the occupation number with the infinite time. Note that $\Omega_f R_H$ is a dimensionless quantity, and we can set $R_H=1$ for simplicity. In Figs.~1(a) and 1(b), the black and the grey curves are for the Hayward black hole with $\ell=0.2$ and $\ell=0.3$, respectively, and the dashed curves are for the Schwarzschild black hole where $\ell=0$.
Thus, in Fig.~1(a),
the low-lying black dot-dashed curve ($\ell=0.5$) and the grey dashed curve ($\ell=0.55$)
are for the occupation number of the Hayward black hole close to the extremal limit $(\ell = \ell_{\max} \approx 0.577)$.}
\end{figure}

Likewise, the spectrum of the occupation number in the Hayward black hole
can also be calculated from Eqs.~\eqref{eq:prob amp} and \eqref{eq:occ num def} as
    \begin{equation}
    \label{eq:occ num2}
    N\left( \eta_f,\omega_f \right) =\left. \frac 1 4 \omega_f \zeta^2 \left[\left(1-\frac{1}{\omega_f \zeta^2}\right)^2
    +\frac{1}{\omega_f^2 \zeta^2} \left(\frac{\textrm d \zeta}{\textrm d \eta}\right)^2
    \right] \right|_{\omega_0 = \sqrt{1-H\eta_f}\omega_f}.
    \end{equation}
By using Eq.~\eqref{eq:time parameter},
a frequency for the original time $t$ can be defined as
$\Omega(t) = (\textrm d \eta/\textrm d t) \omega (\eta) = f(R) \omega (\eta)$,
so the occupation number \eqref{eq:occ num2} can be written in terms of $t_f$ and $\Omega_f$ as \cite{Kolopanis:2013sty}
    \begin{equation}
    \label{eq:finite time}
    N(t_f,\Omega_f) =\frac 1 4 e^{\frac 1 2 Ht_f} \left(\xi^2+\chi^2\right) \left[
    \left(1-\frac{1}{e^{\frac 1 2 Ht_f} (\xi^2+\chi^2)}\right)^2
    +\left(\frac{H e^{-Ht_f}}{\Omega_f} \frac{\xi \dot{\xi}+\chi\dot{\chi}}{\xi^2+\chi^2} \right)^2
    \right],
    \end{equation}
where $\Omega_f \approx e^{-H t_f} \omega_f$ with $\dot \xi = \textrm d \xi / \textrm d \tilde \eta$ and $\dot \chi = \textrm d \chi / \textrm d \tilde \eta$.

On the other hand,  in order to study the limiting case we rewrite the occupation number \eqref{eq:finite time} in terms of $x$ and $y$ for convenience as
\begin{equation}
    \label{eq:eq for xy}
    N(x,y)=\frac{x}{4y} \left(\xi^2+\chi^2\right)
    \left(1-\frac{1}{(x/y) (\xi^2+\chi^2)}\right)^2
    +\frac {1}{y^4} \frac{\{x (\xi h_1+\chi h_2 )\}^2}{(x/y)(\xi^2+\chi^2)},
\end{equation}
where
\begin{align}
    h_1 (x,y) &= e^{-Ht_f} \dot \xi = -\frac{\pi y^2}{4} \{Y_0 (x)J_0 (y)-J_0 (x) Y_0 (y)\},\\
    h_2 (x,y) &=e^{-Ht_f} \dot \chi = -\frac{\pi y^2}{4} \{Y_1 (x)J_0 (y)-J_1 (x) Y_0 (y)\}.
\end{align}
In the incipient limit, $x$ and $y$ are approximately  $x\approx2\Omega_f e^{H t_f/2}/H$ and $y\approx 2\Omega_f/H$, respectively.
Using the asymptotic forms of Bessel functions in the limit $t_f \to\infty$, we get
\begin{align}
    \label{eq:limit_inf1}
    \frac x y (\xi^2 +\chi^2) &\to \frac{\pi y}{2}\{J_1^2 (y)+Y_1^2 (y)\},\\\label{eq:limit_inf2}
    x\left(\xi h_1+\chi h_2\right) &\to -\frac{\pi y^3}{4} \{J_0 (y) J_1 (y)+Y_0 (y) Y_1 (y)\}.
\end{align}
From Eqs.~\eqref{eq:limit_inf1} and \eqref{eq:limit_inf2}, the occupation number \eqref{eq:eq for xy} for $t_f \to\infty$ is obtained as
    \begin{equation}
    \label{eq:infinite time}
    N_\infty (\Omega_f)
    = \frac{\pi}{8} \frac{y}{J_1^2 (y)+Y_1^2 (y)}
    \left[\left(\!\! J_1^2 (y)+Y_1^2 (y)\! -\! \frac{2}{\pi y} \right)^2
    +\{J_0 (y) J_1 (y) +Y_0 (y) Y_1 (y)\}^2 \right].
    \end{equation}

The spectrum of the occupation number in terms of $\Omega_f$  for a finite time
and the infinite time are shown in Fig.~\ref{fig:a} and Fig.~\ref{fig:b}.
The radiation at early times in Fig.~\ref{fig:a} turns out to be nonthermal;
however, the spectrum in Fig.~\ref{fig:b} is very close to the thermal Hawking distribution if the shell approaches its own horizon.
In particular, in Fig.~\ref{fig:a},
we can find that the number of quanta at a low frequency decreases when $\ell$ is getting larger.
Extremely, if $\ell \to \ell_{\max} = 4 M/(3\sqrt 3)$,
the occupation number \eqref{eq:finite time} eventually vanishes because $H=(R_H^2-3\ell^2)/R_H^3$ approaches zero with $R_H=1$.
We can show easily $N(x,y)\to 0$ by using the following limits:
\begin{equation}
    (x/y)(\xi^2+\chi^2) \to 1,\qquad(x/y^2) (\xi h_1+\chi h_2)\to 0,
\end{equation}
which are obtained from the asymptotic forms of Bessel functions as $H\to 0$, i.e., $x\to\infty$ and $y\to\infty$.
The plotting for this result can also be found in Fig.~\ref{fig:a};
the occupation number is getting smaller
as $\ell$ is getting closer to $\ell_{\max} = 1/\sqrt 3 \approx 0.577$ ($R_H =1$).
This extremal limit looks similar to the fact that the Hawking radiation vanishes for extremal black holes.

Our results are quite similar to those of the Reissner-Nordstr\"om (RN) domain wall in Ref.~\cite{Greenwood:2009pd}.
The author in Ref.~\cite{Greenwood:2009pd} showed that the spectrum of the occupation number is nonthermal, but it becomes more and more thermal as $t_f \to\infty$.
In addition, in the extremal case
the temperature seen by the asymptotic observer goes to zero when the shell crosses the horizon.
These similarities are due to the fact that the spacetime structure of Eq.~\eqref{eq:exterior} is analogous to that of the RN domain wall
in that they share two horizons and asymptotic flatnesses.
However, there are two differences between our analysis and that of the RN domain wall:
the methodology and the type of a scalar field.
Firstly, the author in Ref.~\cite{Greenwood:2009pd} found ``best fit temperatures" $\beta^{-1}$
for a finite time
by fitting $\beta\omega_f$ to $\log{(1+1/N)}$,
whereas we investigated the occupation number analytically for infinite time as well as a finite time.
Thus, we could read off the Hawking temperature of the Hayward black hole at infinite time analytically.
Secondly, a complex scalar field which represents charged particles and antiparticles was considered
in order to investigate the effect of charge $Q(Q>0)$ of the RN black hole.
The occupation number for the particles (or antiparticles) with a negative charge
was shown to be subdominant to that with a positive charge because of the Coulomb repulsion.
However, we considered the real scalar field $\Phi$ for neutral particles,
so there exists one kind of the occupation number.
Moreover, instead of $Q$, the effect of $\ell$ appears
as shown in Fig.~\ref{fig:a} and Fig.~\ref{fig:b}.
Essentially, the above differences came from the fact that
the Hayward black hole is neutral unlike the RN black hole in spite of similar spacetime structures.

Now, we can read off
a temperature in a low frequency region.
In the low frequency $\Omega \ll 1$, Eq.~\eqref{eq:infinite time} is approximated as
    \begin{equation}
    \label{eq:low frequency}
    N_\infty (\Omega_f) \approx \frac{H}{4\pi \Omega_f},
    \end{equation}
and the Planckian distribution can also be written as
    \begin{equation}
    \label{eq:Planck dis approximation}
    N_{\textrm{Planck}}(\Omega)=\frac{1}{e^\frac{\Omega}{T}-1}\approx \frac{T}{\Omega}.
    \end{equation}
Comparing Eq.~\eqref{eq:low frequency} and Eq. \eqref{eq:Planck dis approximation},
we can obtain the temperature
    \begin{equation}
    T = \frac{H}{4\pi}
    \end{equation}
which is the same as the Hawking temperature of the Hayward black hole.

\begin{figure}[t]
\centering
\includegraphics[width=0.49\textwidth]{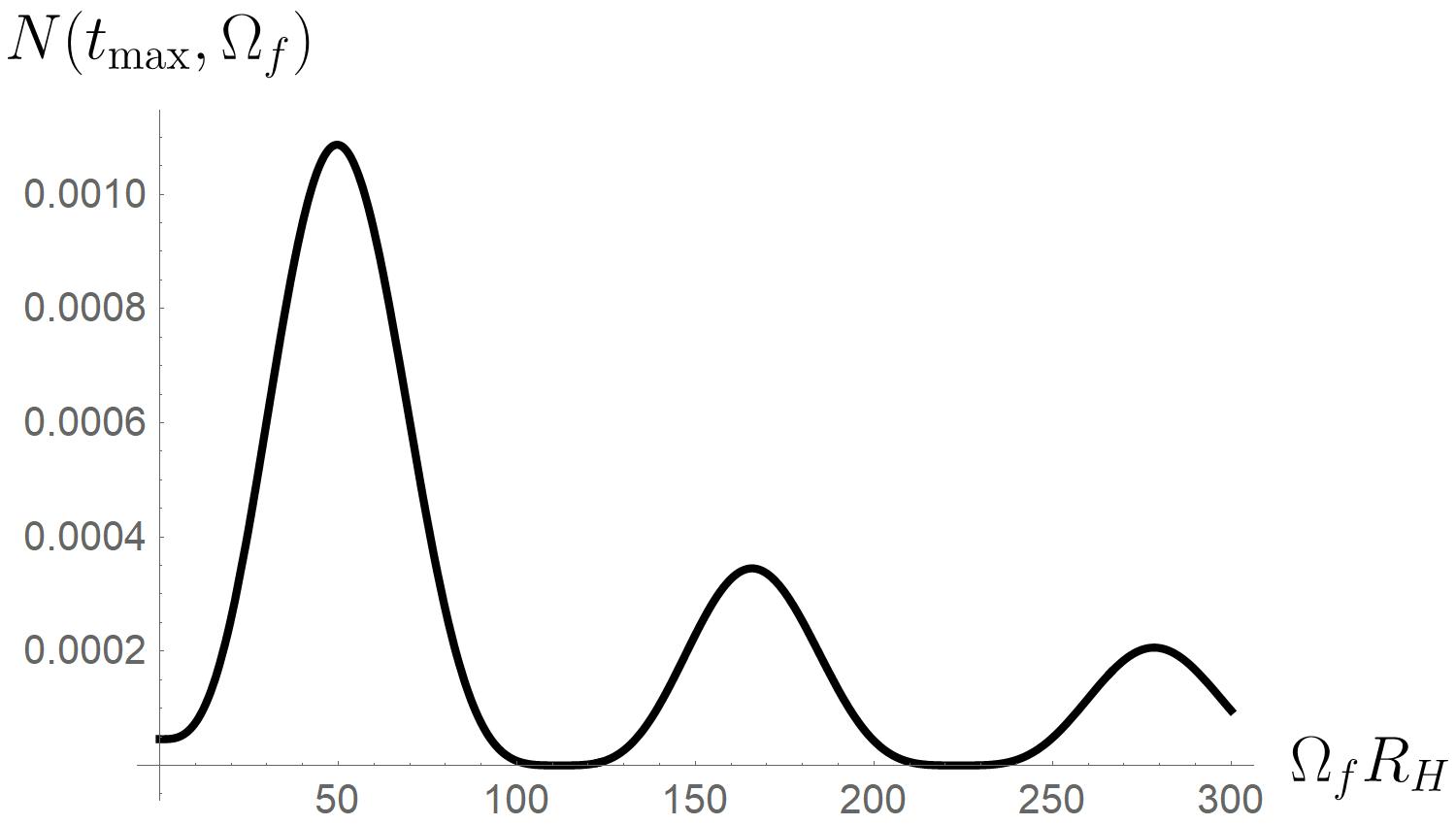}
\caption{The occupation number \eqref{eq:occ num no horizon} for the final time $t_{\max} \approx 0.028$ ($R_0 =0.02$) is plotted with $\tilde H \approx 0.95$, $\ell=0.6$, and $R_H =1$.
The occupation number is damped oscillating, which is different from the thermal Hawking distribution.}
\label{fig:occ num no horizon}
\end{figure}

On the other hand, in the case of $M<M^*$, the occupation number can also be calculated
from the analytic solution \eqref{eq:sol in M<M*},
which is given as
    \begin{equation}
    \label{eq:occ num no horizon}
    N\left( t_f,\Omega_f \right) =\left. \frac 1 4 \Omega_f \bar \zeta^2 \left[\left(1-\frac{1}{\Omega_f \bar \zeta^2}\right)^2
    +\frac{1}{\Omega_f^2 \bar \zeta^2} \left(\frac{\textrm d \bar \zeta}{\textrm d t}\right)^2
    \right] \right|_{\Omega_0 = \sqrt{1-\tilde H t_f}\Omega_f},
    \end{equation}
where $\Omega_0 =\sqrt{\tilde b_0}$ and $\Omega_{f}=\Omega_0/\sqrt{1-\tilde H t}$ with $\tilde H = \tilde b_1 / \tilde b_0$.
Thus, the spectrum of the occupation number in terms of $\Omega_f$ for the final time $t_{\max} = \sqrt 2 R_0$
is plotted in Fig.~\ref{fig:occ num no horizon}.
Note that it takes a finite time during the collapsing and
the spectrum even for the final time $t_{\max}$ is not close to the thermal Hawking distribution
in contrast to the spectrum in Fig.~\ref{fig:b}.

\section{conclusion}
\label{sec:conclusion}
We studied the collapsing shell to form the Hayward black hole and investigated the quantum radiation.
Using the Israel formulation, we obtained the mass relation between
the energy density of the shell and the mass of the Hayward black hole.
The main difference from the Schwarzschild black hole is that
the energy density in the exterior region of the shell as well as the energy density on
the shell contributes to the source of the Hayward black hole.
Next, we investigated the quantum radiation from the shell
by employing the functional Schr\"odinger equation
in cases of $M>M^*$ and $M<M^*$.
In $M>M^*$ when the shell forms the Hayward black hole, the equation
takes the form of
the harmonic oscillator with time-dependent frequency.
From the vacuum state defined by the coherent states,
the density matrix, the probability current, and the occupation number were exactly calculated.
By using these quantities, we showed that the process of the radiation is unitary during the collapsing.
For $M<M^*$,
we derived the functional Schr\"odinger equation for a small radius
which corresponds to the incipient limit in the former case ($M>M^*$).
In this case, all terms in the exterior and interior actions
appear in the quantized Hamiltonian
and then, the functional Schr\"odinger equation is given by the form of the harmonic oscillator
with the time-dependent mass and frequency.
For a certain small time satisfying some constraints,
we could find the analytic solution of the equation.
Next, we investigated the spectrum of the radiation in the case of $M>M^*$ when the shell finally forms the Hayward black hole.
The spectrum
does not coincide with the thermal Hawking radiation at early times; however,
it is very close to the thermal Hawking distribution when the shell approaches the horizon.
For the infinite time,
the temperature of the radiation could be estimated in the limit of the low frequency,
and it turned out to be the Hawking temperature of the Hayward black hole.
In addition, the number of quanta at a low frequency decreases for a larger value of $\ell$.
In the extremal limit of $\ell \to \ell_{\max}$,
the occupation number \eqref{eq:finite time} eventually vanishes, which
is reminiscent of the extremal limit of black holes where the Hawking radiation vanishes.
Moreover, the occupation number for $M<M^*$ was studied, and thus, it turned out to be nonthermal even for the final time when the shell approaches the origin.

Finally, we comment on the incipient limit $R\to R_H$ employed
in our paper.
In the incipient limit, we would get the simplified
analytic results such as Eqs.~\eqref{eq:action approx} and \eqref{eq:solution1},
so we could easily show
that the whole process of the collapsing must be unitary
because $\operatorname{Tr}\rho =1$ regardless of $\eta$ in Eq.~\eqref{eq:unitarity}.
However, the incipient limit corresponds to the moment when the shell approaches the horizon, that is, almost the final stage of the collapsing.
If the incipient limit is released, then the vacuum state \eqref{eq:solution} will be modified.
Although we expect that this modification of the vacuum state will not affect the final results in our paper,
it deserves further study.

\acknowledgments
We would like to thank Myungseok Eune and Yongwan Gim for exciting discussions.
This work was supported by
the National Research Foundation of Korea(NRF) grant funded by the
Korea government(MSIP) (Grant No. 2017R1A2B2006159).


\bibliographystyle{JHEP}       

\bibliography{references}

\providecommand{\href}[2]{#2}\begingroup\raggedright\begin{thebibliography}{10}

\bibitem{Bardeen:1968}
J.~Bardeen{\emph{,~Proceedings of GR5, Tiflis, USSR} (1968) 174}.

\bibitem{Borde:1996df}
A.~Borde, \emph{{Regular black holes and topology change}},
  \href{http://dx.doi.org/10.1103/PhysRevD.55.7615}{\emph{Phys. Rev.} {\bf D55}
  (1997) 7615--7617}, [\href{http://arxiv.org/abs/gr-qc/9612057}{{\tt
  gr-qc/9612057}}].

\bibitem{AyonBeato:1999ec}
E.~Ayon-Beato and A.~Garcia, \emph{{Nonsingular charged black hole solution for
  nonlinear source}},
  \href{http://dx.doi.org/10.1023/A:1026640911319}{\emph{Gen. Rel. Grav.} {\bf
  31} (1999) 629--633}, [\href{http://arxiv.org/abs/gr-qc/9911084}{{\tt
  gr-qc/9911084}}].

\bibitem{Hayward:2005gi}
S.~A. Hayward, \emph{{Formation and evaporation of regular black holes}},
  \href{http://dx.doi.org/10.1103/PhysRevLett.96.031103}{\emph{Phys. Rev.
  Lett.} {\bf 96} (2006) 031103},
  [\href{http://arxiv.org/abs/gr-qc/0506126}{{\tt gr-qc/0506126}}].

\bibitem{Bambi:2013ufa}
C.~Bambi and L.~Modesto, \emph{{Rotating regular black holes}},
  \href{http://dx.doi.org/10.1016/j.physletb.2013.03.025}{\emph{Phys. Lett.}
  {\bf B721} (2013) 329--334}, [\href{http://arxiv.org/abs/1302.6075}{{\tt
  1302.6075}}].

\bibitem{Balart:2014cga}
L.~Balart and E.~C. Vagenas, \emph{{Regular black holes with a nonlinear
  electrodynamics source}},
  \href{http://dx.doi.org/10.1103/PhysRevD.90.124045}{\emph{Phys. Rev.} {\bf
  D90} (2014) 124045}, [\href{http://arxiv.org/abs/1408.0306}{{\tt
  1408.0306}}].

\bibitem{Halilsoy:2013iza}
M.~Halilsoy, A.~Ovgun and S.~H. Mazharimousavi, \emph{{Thin-shell wormholes
  from the regular Hayward black hole}},
  \href{http://dx.doi.org/10.1140/epjc/s10052-014-2796-4}{\emph{Eur. Phys. J.}
  {\bf C74} (2014) 2796}, [\href{http://arxiv.org/abs/1312.6665}{{\tt
  1312.6665}}].

\bibitem{Abbas:2014oua}
G.~Abbas and U.~Sabiullah, \emph{{Geodesic Study of Regular Hayward Black
  Hole}}, \href{http://dx.doi.org/10.1007/s10509-014-1992-x}{\emph{Astrophys.
  Space Sci.} {\bf 352} (2014) 769--774},
  [\href{http://arxiv.org/abs/1406.0840}{{\tt 1406.0840}}].

\bibitem{Amir:2015pja}
M.~Amir and S.~G. Ghosh, \emph{{Rotating Hayward’s regular black hole as
  particle accelerator}},
  \href{http://dx.doi.org/10.1007/JHEP07(2015)015}{\emph{JHEP} {\bf 07} (2015)
  015}, [\href{http://arxiv.org/abs/1503.08553}{{\tt 1503.08553}}].

\bibitem{Pourhassan:2016qoz}
B.~Pourhassan, M.~Faizal and U.~Debnath, \emph{{Effects of Thermal Fluctuations
  on the Thermodynamics of Modified Hayward Black Hole}},
  \href{http://dx.doi.org/10.1140/epjc/s10052-016-3998-8}{\emph{Eur. Phys. J.}
  {\bf C76} (2016) 145}, [\href{http://arxiv.org/abs/1603.01457}{{\tt
  1603.01457}}].

\bibitem{Chiba:2017nml}
T.~Chiba and M.~Kimura, \emph{{A note on geodesics in the Hayward metric}},
  \href{http://dx.doi.org/10.1093/ptep/ptx037}{\emph{PTEP} {\bf 2017} (2017)
  043E01}, [\href{http://arxiv.org/abs/1701.04910}{{\tt 1701.04910}}].

\bibitem{Mehdipour:2016vxh}
S.~H. Mehdipour and M.~H. Ahmadi, \emph{{Black Hole Remnants in Hayward
  Solutions and Noncommutative Effects}},
  \href{http://dx.doi.org/10.1016/j.nuclphysb.2017.09.021}{\emph{Nucl. Phys.}
  {\bf B926} (2018) 49--69}, [\href{http://arxiv.org/abs/1604.08584}{{\tt
  1604.08584}}].

\bibitem{Vachaspati:2006ki}
T.~Vachaspati, D.~Stojkovic and L.~M. Krauss, \emph{{Observation of incipient
  black holes and the information loss problem}},
  \href{http://dx.doi.org/10.1103/PhysRevD.76.024005}{\emph{Phys. Rev.} {\bf
  D76} (2007) 024005}, [\href{http://arxiv.org/abs/gr-qc/0609024}{{\tt
  gr-qc/0609024}}].

\bibitem{Vachaspati:2007hr}
T.~Vachaspati and D.~Stojkovic, \emph{{Quantum radiation from quantum
  gravitational collapse}},
  \href{http://dx.doi.org/10.1016/j.physletb.2008.04.004}{\emph{Phys. Lett.}
  {\bf B663} (2008) 107--110}, [\href{http://arxiv.org/abs/gr-qc/0701096}{{\tt
  gr-qc/0701096}}].

\bibitem{Greenwood:2008zg}
E.~Greenwood and D.~Stojkovic, \emph{{Hawking radiation as seen by an infalling
  observer}},
  \href{http://dx.doi.org/10.1088/1126-6708/2009/09/058}{\emph{JHEP} {\bf 09}
  (2009) 058}, [\href{http://arxiv.org/abs/0806.0628}{{\tt 0806.0628}}].

\bibitem{Greenwood:2009gp}
E.~Greenwood, E.~Halstead and P.~Hao, \emph{{Classical and Quantum Equations of
  Motion for a BTZ Black String in AdS Space}},
  \href{http://dx.doi.org/10.1007/JHEP02(2010)044}{\emph{JHEP} {\bf 02} (2010)
  044}, [\href{http://arxiv.org/abs/0912.1860}{{\tt 0912.1860}}].

\bibitem{Greenwood:2009pd}
E.~Greenwood, \emph{{Hawking Radiation from a Reisner-Nordstrom Domain Wall}},
  \href{http://dx.doi.org/10.1088/1475-7516/2010/01/002}{\emph{JCAP} {\bf 1001}
  (2010) 002}, [\href{http://arxiv.org/abs/0910.0024}{{\tt 0910.0024}}].

\bibitem{Greenwood:2010sx}
E.~Greenwood, D.~I. Podolsky and G.~D. Starkman, \emph{{Pre-Hawking Radiation
  from a Collapsing Shell}},
  \href{http://dx.doi.org/10.1088/1475-7516/2011/11/024}{\emph{JCAP} {\bf 1111}
  (2011) 024}, [\href{http://arxiv.org/abs/1011.2219}{{\tt 1011.2219}}].

\bibitem{Kolopanis:2013sty}
M.~Kolopanis and T.~Vachaspati, \emph{{Quantum Excitations in Time-Dependent
  Backgrounds}},
  \href{http://dx.doi.org/10.1103/PhysRevD.87.085041}{\emph{Phys. Rev.} {\bf
  D87} (2013) 085041}, [\href{http://arxiv.org/abs/1302.1449}{{\tt
  1302.1449}}].

\bibitem{Saini:2015dea}
A.~Saini and D.~Stojkovic, \emph{{Radiation from a collapsing object is
  manifestly unitary}},
  \href{http://dx.doi.org/10.1103/PhysRevLett.114.111301}{\emph{Phys. Rev.
  Lett.} {\bf 114} (2015) 111301}, [\href{http://arxiv.org/abs/1503.01487}{{\tt
  1503.01487}}].

\bibitem{Saini:2016rrt}
A.~Saini and D.~Stojkovic, \emph{{Hawking-like radiation and the density matrix
  for an infalling observer during gravitational collapse}},
  \href{http://dx.doi.org/10.1103/PhysRevD.94.064028}{\emph{Phys. Rev.} {\bf
  D94} (2016) 064028}, [\href{http://arxiv.org/abs/1609.06584}{{\tt
  1609.06584}}].

\bibitem{Saini:2017mur}
A.~Saini and D.~Stojkovic, \emph{{Gravitational collapse and Hawking-like
  radiation of a shell in AdS spacetime}},
  \href{http://dx.doi.org/10.1103/PhysRevD.97.025020}{\emph{Phys. Rev.} {\bf
  D97} (2018) 025020}, [\href{http://arxiv.org/abs/1711.08182}{{\tt
  1711.08182}}].

\bibitem{Das:2019iru}
A.~Das and N.~Banerjee, \emph{{Unitarity in Reissner–Nordström background:
  striding away from information loss}},
  \href{http://dx.doi.org/10.1140/epjc/s10052-019-6991-1}{\emph{Eur. Phys. J.}
  {\bf C79} (2019) 475}, [\href{http://arxiv.org/abs/1902.03378}{{\tt
  1902.03378}}].

\bibitem{Israel:1966rt}
W.~Israel, \emph{{Singular hypersurfaces and thin shells in general
  relativity}}, \href{http://dx.doi.org/10.1007/BF02710419,
  10.1007/BF02712210}{\emph{Nuovo Cim.} {\bf B44S10} (1966) 1}.

\bibitem{Ipser:1983db}
J.~Ipser and P.~Sikivie, \emph{{The Gravitationally Repulsive Domain Wall}},
  \href{http://dx.doi.org/10.1103/PhysRevD.30.712}{\emph{Phys. Rev.} {\bf D30}
  (1984) 712}.

\bibitem{Lopez:1988gt}
C.~A. Lopez, \emph{{Dynamics of Charged Bubbles in General Relativity and
  Models of Particles}},
  \href{http://dx.doi.org/10.1103/PhysRevD.38.3662}{\emph{Phys. Rev.} {\bf D38}
  (1988) 3662--3666}.

\bibitem{goldstein:mechanics}
H.~Goldstein, \emph{Classical Mechanics}.
\newblock Addison-Wesley, 1980.

\bibitem{Dantas:1990rk}
C.~M.~A. Dantas, I.~A. Pedrosa and B.~Baseia, \emph{{Harmonic oscillator with
  time-dependent mass and frequency and a perturbative potential}},
  \href{http://dx.doi.org/10.1103/PhysRevA.45.1320}{\emph{Phys. Rev.} {\bf A45}
  (1992) 1320--1324}.

\end{thebibliography}\endgroup

\end{document}